\documentclass[11pt,twoside]{article}
\usepackage{graphicx,epsfig,natbib,epstopdf,amssymb}
\usepackage{CS18}
\begin{document}
%
%
%
\title{Hot Stars With Cool Companions}
%
%
\author{Kevin Gullikson$^{1}$, Adam Kraus$^{1}$, Sarah Dodson-Robinson$^{2}$}
\affil{$^1$University of Texas, 2515 Speedway, STOP C1400, Austin, TX 78712}
\affil{$^2$University of Delaware, 217 Sharp Lab, Newark, DE 19716}
\begin{abstract}
%
%

Young intermediate-mass stars have become high-priority targets for direct-imaging planet searches following the recent discoveries of planets orbiting e.g. HR 8799 and Beta Pictoris. Close stellar companions to these stars can affect the formation and orbital evolution of any planets, and so a census of the multiplicity properties of nearby intermediate mass stars is needed. Additionally, the multiplicity can help constrain the important binary star formation physics. We report initial results from a spectroscopic survey of 400 nearby A- and B-type stars. We search for companions by cross-correlating high resolution and high signal-to-noise ratio echelle spectra of the targets stars against model spectra for F- to M-type stars. We have so far found 18 new candidate companions, and have detected the spectral lines of the secondary in 4 known spectroscopic binary systems. We present the distribution of mass-ratios for close companions, and find that it differs from the distribution for wide ($a \gtrsim 100$ AU) intermediate-mass binaries, which may indicate a different formation mechanism for the two populations.

\end{abstract}
%
%
%
%
%
\section{Introduction}

Recent radial-velocity searches for massive planets orbiting $1 - 2 M_{\odot}$ subgiant ``retired A-stars'' \citep{Johnson2011} have indicated that intermediate-mass stars may be more likely to host planets than solar-type and low-mass stars. This finding, in combination with the direct-imaging detections of massive planets on very wide orbits around young A-type stars \cite[e.g.][]{Marois2008, Marois2010, Lagrange2010}, has spurred an increased interest in intermediate-mass stars as potential planet hosts. Especially interesting are the \emph{young} intermediate-mass stars, where a planet on a wide orbit will be bright enough to be directly imaged and characterized. However, since stellar multiplicity increases with mass \citep{Zinnecker2007} and decreases with age \citep{Duchene2013}, the same young intermediate-mass stars that are attractive targets for planet searches may very often host a close \emph{stellar} companion that can impact the planet formation process. Since close companions that may not be resolvable by current imaging systems are likely to have the largest impact on planet formation, a spectroscopic multiplicity survey of nearby intermediate-mass stars is needed.

The multiplicity of A- and B-type stars can also help constrain the binary star formation mechanism and the relevant physics involved during and after the formation of the secondary. The dominant mode of binary star formation is thought to be molecular core fragmentation \citep{Boss1979, Boss1986, Bate1995} in which a collapsing cloud of gas fragments into two or more stars. The ratio of masses is set largely by the turbulent power spectrum, density structure, angular momentum, and magnetic field in the pre-stellar core. Low-mass stars may also form directly via gravitational instabilities \citep{Kratter2006, Stamatellos2011} in the massive disk surrounding a forming intermediate-mass star. This alternate formation scenario will only act within $\sim 100$ AU, and may create an observationally distinguishable inner mass-ratio distribution. Observational studies of the multiplicity, mass-ratio distribution and separation distribution in a population of stars provide a benchmark against which future star formation models and simulations must agree.

The intermediate-mass binary population has been well-mapped for wide orbits ($a \gtrsim 50$ AU) with imaging studies in the young Scorpius-Centaurus OB associations \citep{Kouwenhoven2007} and recently for field A-star primaries \citep{DeRosa2014}. These studies have both found a preference for binary systems with low mass-ratios q ($q \equiv M_s / M_p$, where $M_s$ and $M_p$ are the secondary and primary mass, respectively), and are consistent with a power law $f(q) \sim q^{\Gamma}$ with $\Gamma = -0.4$. While there does not appear to be a difference between close and wide binary systems for solar-type and low-mass primaries \citep{Meyer2013}, there is some evidence that intermediate-mass binaries have a flatter mass-ratio distribution inside $\sim 100$ AU  \citep{DeRosa2014}. This radius is the same order of magnitude as a circumstellar disk, and seems to imply that the disk around an intermediate-mass protostar plays a more crucial role in binary formation and evolution than it does for solar-type and low-mass stars. It may be a sign that disk fragmentation is capable of producing a distinct population of binary companions when there is enough disk mass, or could be an effect of increased or preferential accretion onto the secondary star if it forms near the gas-rich disk.

Most of the systematic studies searching for companions to intermediate-mass stars to date have used imaging, and so miss low-mass companions within a few tens of AU from the primary star. Thus, a spectroscopic study is necessary to derive the true mass-ratio distribution for close binary systems and compare it to that of wide binary systems. We have begun such a survey, using high signal-to-noise ratio (S/N) echelle spectra to directly detect the spectral lines of secondary stars orbiting 400 nearby Main Sequence A- and B-type stars.

\section{Observations and Methods}
\label{sec:obs}

We have observed 292 of a total sample of 400 A- and B-type stars. The sample was chosen from the Simbad database, and includes all main-sequence A- and B-stars with $v\sin{i} > 80$ km $s^{-1}$, $m_V < 6$, and no spectral peculiarities. We have observed these stars with the CHIRON spectrograph on the 1.5m telescope at Cerro Tololo Interamerican Observatory, the Tull coude spectrograph on the 2.7m telescope at McDonald Observatory, and the High Resolution Spectrograph on the Hobby Eberly Telescope at McDonald Observatory. The data were bias-subtracted, flat-fielded, and extracted with the optimal extraction method using standard IRAF tasks. The extracted spectra were wavelength-calibrated using Th-Ar lamps taken the same night as the data. All three instruments are visible echelle spectrographs with similar resolution and wavelength ranges.

After extraction, the data were corrected for telluric absorption lines using the TelFit package \citep{Gullikson2014}. Several frames were taken for each star, and each frame was telluric-corrected separately to better account for the changing airmass and atmospheric conditions over the course of the exposures. The corrected frames for each target were added together before further analysis, resulting in spectra with a typical peak S/N ratio per pixel of 500. 

As a final pre-processing step, we removed the rotationally broadened spectral lines of the primary star by fitting a cubic Savitzky-Golay \citep{savitzky1964} smoothing spline with a window size of $0.8 v\sin{i}$, where the $v\sin{i}$ came from the most recent literature value in the Simbad database. The factor of 0.8 was chosen to give the best fit to the data while leaving the high frequency components intact. We divided the data by the smoothing spline, in effect passing it through a high-pass filter.

To search for companions, we cross-correlated the processed spectra against the following grid of Phoenix model spectra \citep{Hauschildt2005} 
\begin{itemize}
\item $3000 K < T_{\rm eff} < 7000 K$, in 100 K steps
\item $-0.5 <$ [Fe/H] $< +0.5$, in 0.5 dex steps
\item  $\log{g} = 4.5$
\item $v\sin{i} = 10,20,30,40$ km $s^{-1}$
\end{itemize}

\begin{figure}
\centering
\plottwo{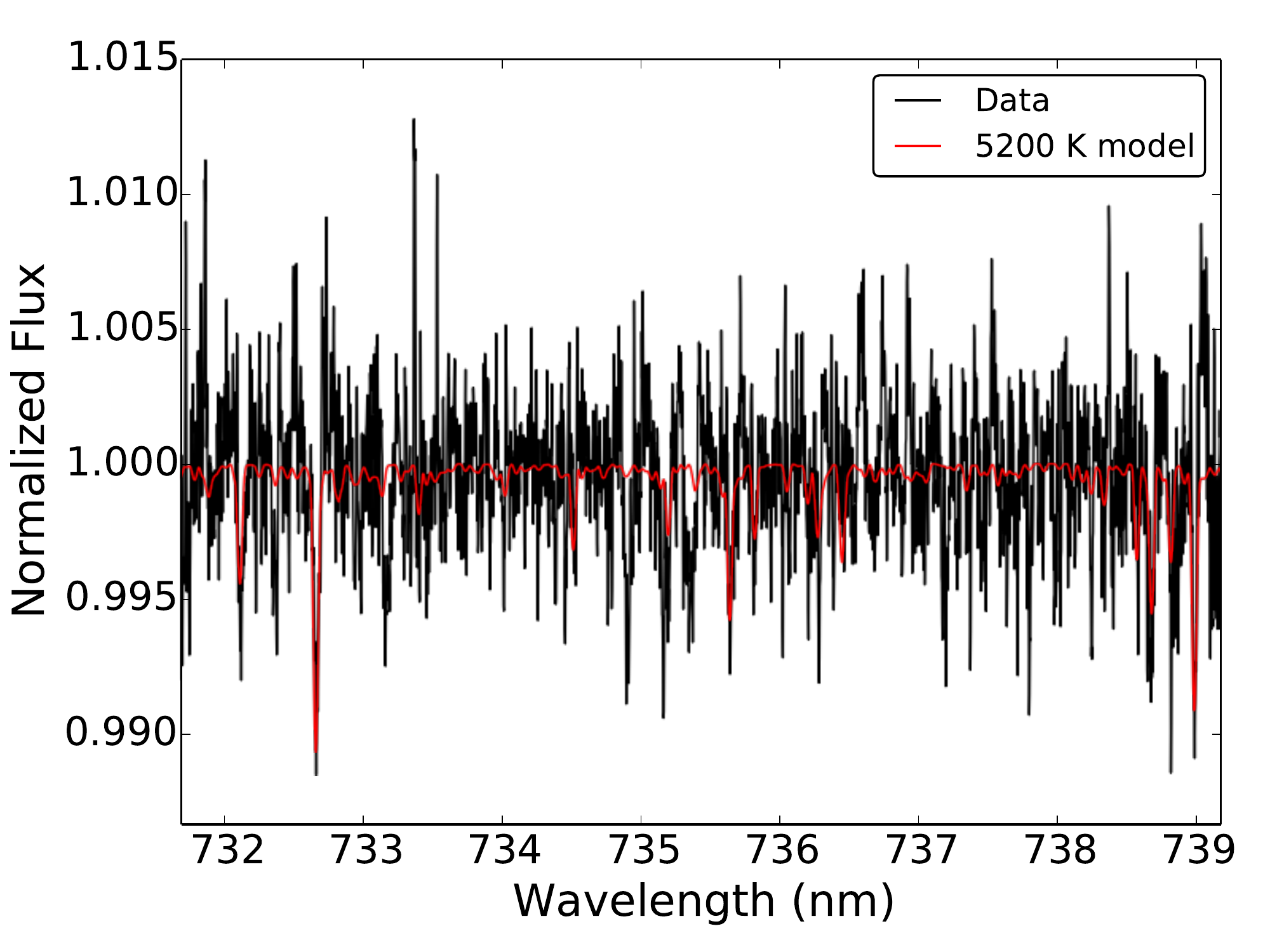}{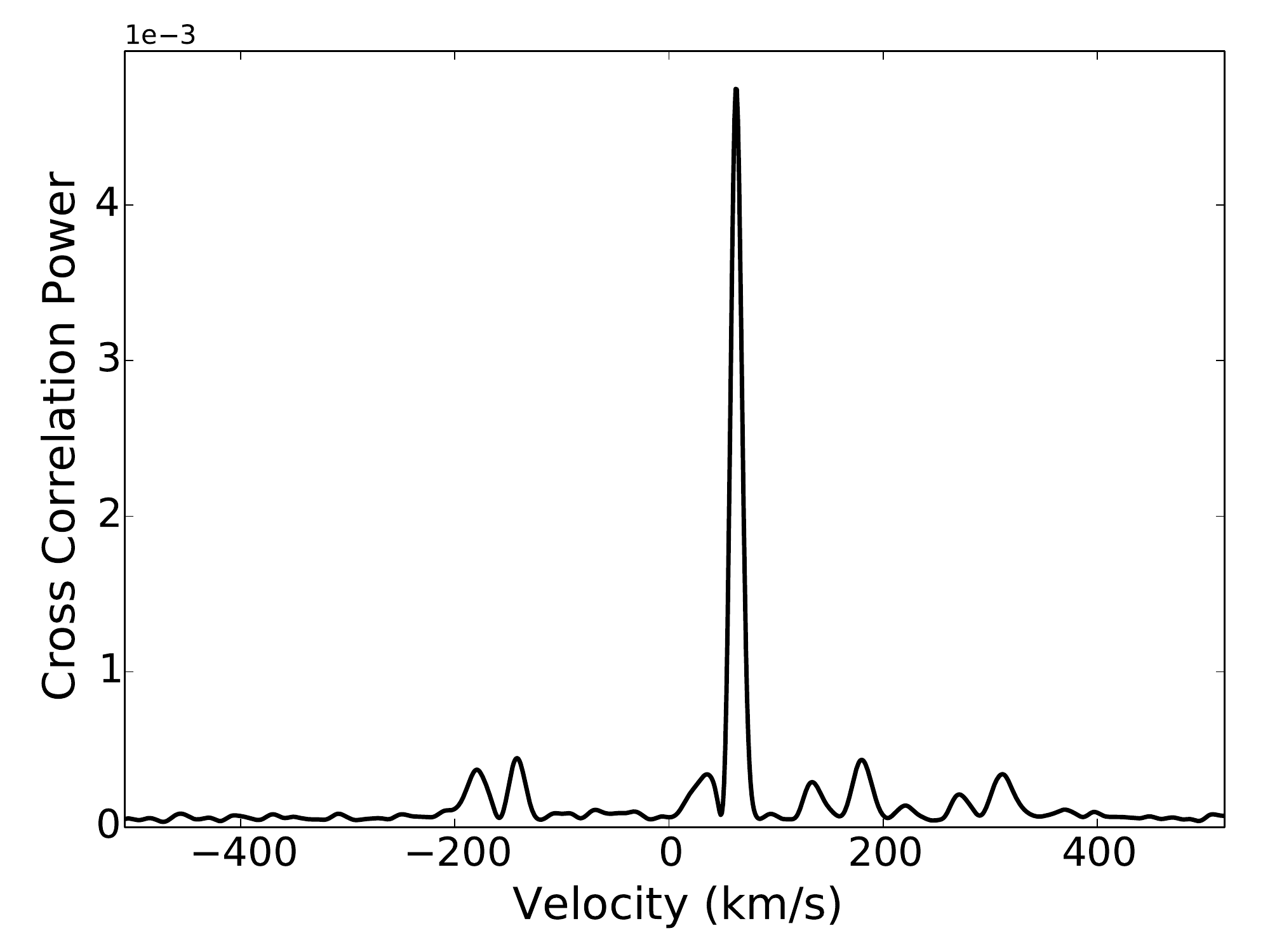} 

\caption{Example of the cross-correlation based technique described in section \ref{sec:obs} \emph{Left:} One order of a spectrum of HIP 32607, an A8V star. The observed spectrum is in black, with a 5200 K model spectrum in red. Some of the spectral lines of the secondary are barely visible in the spectrum, most notably the line near 732.7 nm. \emph{Right:} The cross-correlation function of all orders of the same observation against the same 5200 K model. The strong peak indicates a clear detection of the secondary.}
\label{fig:method}
\end{figure}

Figure \ref{fig:method} demonstrates the cross-correlation method for a secondary star that is just barely detectable ``by eye,'' but is unambiguously detected in the cross-correlation function. This method can detect fainter companions which are not visible in the spectrum but are evident in the cross-correlation function; by injecting artifical signals into the data, we have found that companions with $T_{\rm eff} \gtrsim 3500$ K (spectral types earlier than about M2-M3) are detectable in most of our data.

\section{Preliminary Results and Discussion}
\label{sec:results}

After searching for companions in all of our data to date, we have found 18 new candidate companions, and have detected the spectral lines of the secondary in 4 previously single-lined spectroscopic binaries. Since the new detections require follow-up observations to confirm, we do not report them here. However, we list the companions to known single-lined binaries in Table 1. Since we only have single-epoch data for these stars, we cannot fit an orbital solution. We report the spectral type of the secondary star in Table 1, which is determined from the temperature which gives the most significant cross-correlation function detection.

\begin{table}[h]
  \centering
  \begin{tabular}{l|cc}
    Star Name & Primary Spectral Type & Secondary Spectral Type \\ \hline
    HIP 106786 & A7V & G7 \\
    HIP 32607 & A8V & K0 \\
    HIP 109521 & A5V & K2 \\
    HIP 22833 & A3V & G9 \\
  \end{tabular}
\end{table}

\begin{figure}
\centering
\plotone{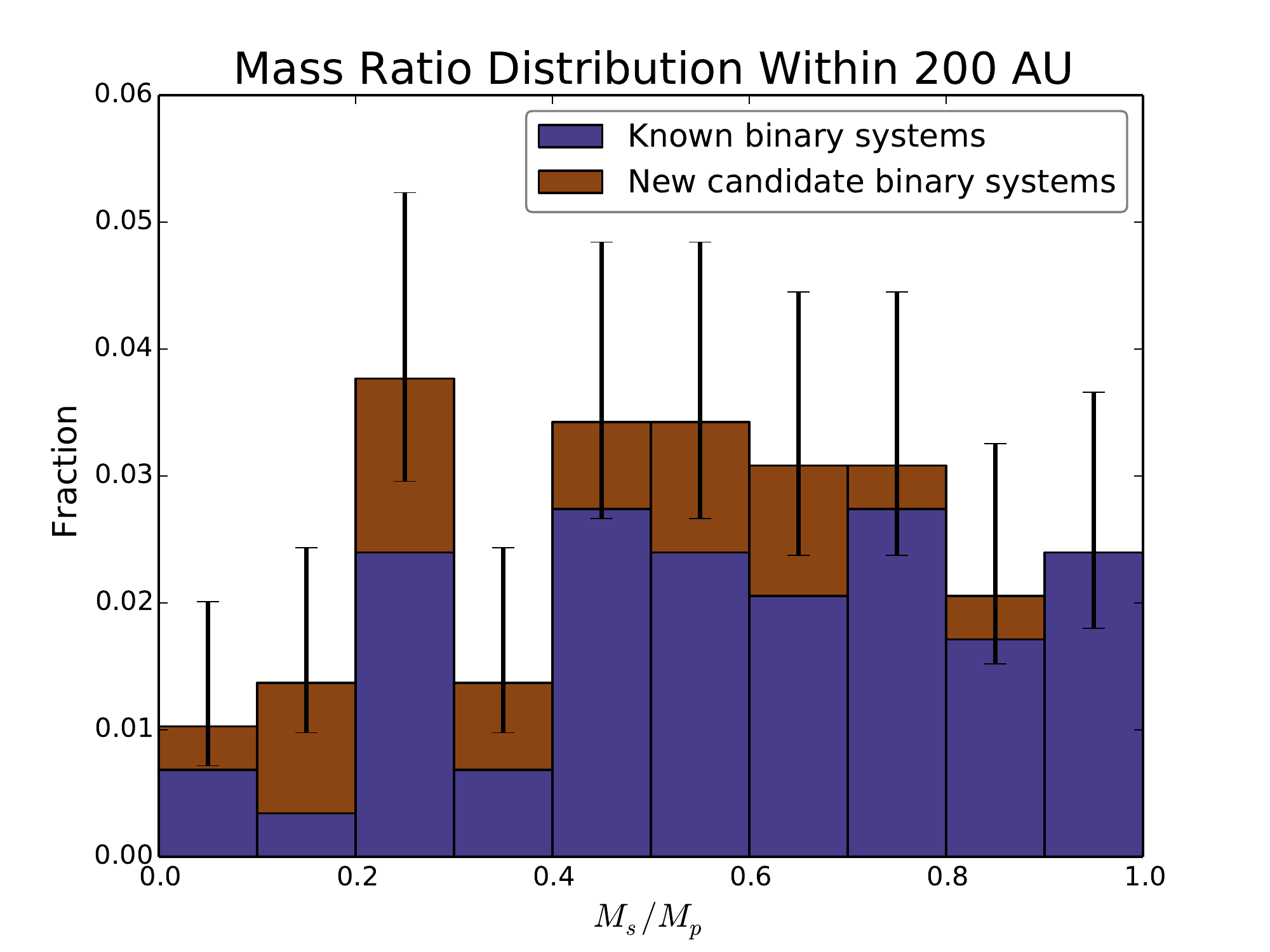}
\caption{Mass-ratio distribution for the stars we have so far observed. The distribution is complete down to $q \sim 0.15$, and includes both known and new binary systems. The $1 \sigma$ binomial confidence interval is shown for each bin.}
\label{fig:massratio}
\end{figure}

We also show the mass-ratio distribution that we derive from the candidates so far detected in Figure \ref{fig:massratio}. To determine the mass of the secondaries, we first find the temperature that gives the most significant cross-correlation function peak. That temperature is the best match to the observed spectrum, and so we take it as the true temperature of the candidate companion star. We then interpolate Baraffe stellar evolutionary tracks \citep{Baraffe1998} at the age of the system to find the secondary star mass. Masses for the primary stars come from main sequence relationships. We do not have ages for most of our sample at this time, and use the main sequence lifetime of the primary star for the system age. Doing so will tend to slightly overestimate the mass of the secondary; however, since we chose only main sequence targets, we don't expect the age to change by much more than a factor of 2 and so the secondary masses should not be significantly affected in most cases. In the near future, we intend to use the observed spectra to measure the effective temperature, gravity, and metallicity of the primary stars, and use that information to better constrain the age of the system and mass of the primary star.

The mass ratio distribution we show in Figure \ref{fig:massratio} is consistent with a flat distribution, similar to the results that \citet{DeRosa2014} find for field A-type stars inside 125 AU. Notably, a Kolmogorov-Smirnov test shows that it is inconsistent with a power law with slope $\Gamma = -0.4$ ($p = 7.4$ x $10^{-6}$). These results seem to indicate that disk physics are important in forming intermediate-mass binaries systems.

\acknowledgments{
This research has made use of the SIMBAD database,
operated at CDS, Strasbourg, France. This project was funded by a UT Austin Hutchinson fellowship to Kevin Gullikson and start-up funding to Sarah Dodson-Robinson from the University of Texas.
}

\normalsize



%
%
%
%
%



\end{document}